\PassOptionsToPackage{unicode}{hyperref}
\PassOptionsToPackage{hyphens}{url}
\PassOptionsToPackage{dvipsnames,svgnames,x11names}{xcolor}
\documentclass[
  12pt]{article}

\usepackage{amsmath,amssymb}
\usepackage{iftex}
\ifPDFTeX
  \usepackage[T1]{fontenc}
  \usepackage[utf8]{inputenc}
  \usepackage{textcomp} 
\else 
  \usepackage{unicode-math}
  \defaultfontfeatures{Scale=MatchLowercase}
  \defaultfontfeatures[\rmfamily]{Ligatures=TeX,Scale=1}
\fi
\usepackage{lmodern}
\ifPDFTeX\else  
\fi
\IfFileExists{upquote.sty}{\usepackage{upquote}}{}
\IfFileExists{microtype.sty}{
  \usepackage[]{microtype}
  \UseMicrotypeSet[protrusion]{basicmath} 
}{}
\makeatletter
\@ifundefined{KOMAClassName}{
  \IfFileExists{parskip.sty}{%
    \usepackage{parskip}
  }{
    \setlength{\parindent}{0pt}
    \setlength{\parskip}{6pt plus 2pt minus 1pt}}
}{
  \KOMAoptions{parskip=half}}
\makeatother
\usepackage{xcolor}
\makeatletter
\ifx\paragraph\undefined\else
  \let\oldparagraph\paragraph
  \renewcommand{\paragraph}{
    \@ifstar
      \xxxParagraphStar
      \xxxParagraphNoStar
  }
  \newcommand{\xxxParagraphStar}[1]{\oldparagraph*{#1}\mbox{}}
  \newcommand{\xxxParagraphNoStar}[1]{\oldparagraph{#1}\mbox{}}
\fi
\ifx\subparagraph\undefined\else
  \let\oldsubparagraph\subparagraph
  \renewcommand{\subparagraph}{
    \@ifstar
      \xxxSubParagraphStar
      \xxxSubParagraphNoStar
  }
  \newcommand{\xxxSubParagraphStar}[1]{\oldsubparagraph*{#1}\mbox{}}
  \newcommand{\xxxSubParagraphNoStar}[1]{\oldsubparagraph{#1}\mbox{}}
\fi
\makeatother

\usepackage{longtable,booktabs,array}
\usepackage{calc} 
\usepackage{etoolbox}
\makeatletter
\patchcmd\longtable{\par}{\if@noskipsec\mbox{}\fi\par}{}{}
\makeatother
\IfFileExists{footnotehyper.sty}{\usepackage{footnotehyper}}{\usepackage{footnote}}
\makesavenoteenv{longtable}
\usepackage{graphicx}
\makeatletter
\def\maxwidth{\ifdim\Gin@nat@width>\linewidth\linewidth\else\Gin@nat@width\fi}
\def\maxheight{\ifdim\Gin@nat@height>\textheight\textheight\else\Gin@nat@height\fi}
\makeatother
\setkeys{Gin}{width=\maxwidth,height=\maxheight,keepaspectratio}
\makeatletter
\def\fps@figure{htbp}
\makeatother

\makeatletter
\@ifpackageloaded{caption}{}{\usepackage{caption}}
\AtBeginDocument{%
\ifdefined\contentsname
  \renewcommand*\contentsname{Table of contents}
\else
  \newcommand\contentsname{Table of contents}
\fi
\ifdefined\listfigurename
  \renewcommand*\listfigurename{List of Figures}
\else
  \newcommand\listfigurename{List of Figures}
\fi
\ifdefined\listtablename
  \renewcommand*\listtablename{List of Tables}
\else
  \newcommand\listtablename{List of Tables}
\fi
\ifdefined\figurename
  \renewcommand*\figurename{Figure}
\else
  \newcommand\figurename{Figure}
\fi
\ifdefined\tablename
  \renewcommand*\tablename{Table}
\else
  \newcommand\tablename{Table}
\fi
}
\@ifpackageloaded{float}{}{\usepackage{float}}
\floatstyle{ruled}
\@ifundefined{c@chapter}{\newfloat{codelisting}{h}{lop}}{\newfloat{codelisting}{h}{lop}[chapter]}
\floatname{codelisting}{Listing}

\makeatother
\makeatletter
\@ifpackageloaded{caption}{}{\usepackage{caption}}
\@ifpackageloaded{subcaption}{}{\usepackage{subcaption}}
\makeatother

\ifLuaTeX
  \usepackage{selnolig}  
\fi
\usepackage[]{natbib}
\usepackage{bookmark}

\IfFileExists{xurl.sty}{\usepackage{xurl}}{} 
\hypersetup{
  pdftitle={Title},
  pdfauthor={Author 1; Author 2},
  pdfkeywords={3 to 6 keywords, that do not appear in the title},
  colorlinks=true,
  linkcolor={blue},
  filecolor={Maroon},
  citecolor={Blue},
  urlcolor={Blue},
  pdfcreator={LaTeX via pandoc}}

\newcommand{\anon}{1}

\usepackage{amsthm}
\newtheorem{definition}{Definition}[section]
\newtheorem{theorem}{Theorem}[section]

\begin{document}

\def\spacingset#1{\renewcommand{\baselinestretch}%
{#1}\small\normalsize} \spacingset{1}


\if1\anon
{
  \title{\bf Rapid Approximation Prediction for Kriging}
  \author{Ziyu Li\thanks{
    \textit{Corresponding author: 
Department of Applied Mathematics and Statistics, 1500 Illinois St, Golden, CO 80401. Email: Ziyu\_Li@mines.edu.}}\hspace{.2cm}, Gregory Fasshauer, and Douglas Nychka\\
	\vspace{0.1cm}\\
    Department of Applied Mathematics and Statistics, \\ Colorado School of Mines. }
  \maketitle
} \fi

\if0\anon
{
  \bigskip
  \bigskip
  \bigskip
  \begin{center}
    {\LARGE\bf Title}
\end{center}
  \medskip
} \fi

\bigskip
\begin{abstract}
Exact Kriging and conditional simulation (CS) for uncertainty quantification are computationally infeasible for modern spatial analyses with large numbers of observations and dense prediction grids. We present a rapid approximation to the Kriging prediction step for stationary Gaussian processes for a regular prediction grid by approximating each off-grid covariance vector by a sparse linear combination of on-grid covariances within a local $L$-order neighborhood of $M = (2L)^2$ neighboring grid points. This reformulation reduces complexity from $\mathcal{O}(N n^3)$ to $\mathcal{O}(N \log N + nM + M^3)$ while preserving accuracy. A factorial study shows that approximation error decreases systematically with increased Mat\'{e}rn smoothness, neighbor order $L$, and grid resolution, aligning with bounds from kernel approximation theory. In a North American summer-rainfall application ($n=1368$), our method produces predictions visually indistinguishable from exact Kriging with point-wise errors on the order of $10^{-5}$ inches and achieves more than $150$ times speedups at a $350\times350$ grid, also outperforming Vecchia and LatticeKrig predictions. Embedded in a fast CS scheme, the approach reproduces Kriging standard errors and scales favorably with both $n$ and $N$. We recommend a practical workflow that uses a fast method for parameter estimation followed by our rapid predictor for fine-grid mapping and uncertainty quantification.\end{abstract}

\noindent%
\vfill

\newpage
\spacingset{1.8} 

\section{Introduction}\label{sec-intro}

A central problem in spatial statistics is the modeling of point-wise scattered or gridded data with spatial components and making predictions at unobserved locations. The field of spatial statistics builds on Tober's 
 First Law of Geography:  
 ``everything is related to everything else, but near things are more related than distant things'' \citep{Tobler1970} using the covariance function of a Gaussian process as a way to quantify relatedness.  In addition to making predictions at unobserved locations, spatial statistical tools can also attach a measure of uncertainty to those predictions. Both of these tasks can be prohibitively time consuming using exact methods with many observations and at many prediction locations. In this work we propose a simple but accurate approximation to the prediction step in a spatial statistical analysis based on stationary Gaussian processes (GP) and document dramatic speedups in the computation.
 
Our analysis is based on the observational spatial model under a universal Kriging context
\begin{equation}\label{eq:ObservationalModel}
z_i = X_i \beta + g(\mathbf{s}_i^O) + \epsilon_i, \text{ for } i = 1, \dots, n.
\end{equation}
We use $z_i$ to denote the $i$-th observation at its possibly irregularly distributed spatial location $\mathbf{s}_i^O$. Note that $\mathbf{s}_i^O$ is the spatial location and the ``$O$'' indicates it is an observation location. $X_i$ are known covariates as part of the linear model fixed process, and $g$ is a mean zero Gaussian process that depends on the spatial location. $\epsilon_i$ are independent errors with $\epsilon_i \sim N(0,\tau^2)$ with $\tau^2$ also known as the {\it nugget variance}. The goal is to draw inferences about the linear model parameters, $\beta$, and the spatial field $f_j = X_j\beta + g(\mathbf{s}_j)$ with $j = 1, \dots, N$, where $j$ indexes unobserved locations.

The base method for estimating $f$ we term {\it Kriging} \citep{Krige1951}, even though we will make more distributional assumptions than in geostatistics. Throughout we denote the spatial estimate as $\hat{f}$. Under the assumptions that the covariance function for $g$ and $\tau^2$ are known, then $\hat{f}$ will be the best linear unbiased estimate for $f$, and the distribution of $\hat{f}$ is multivariate normal with covariance matrix depending on the locations of observations and predictions and $\tau^2$ \citep{Matheron1965}. Despite having a closed form, using this distribution analytically for statistical inference is difficult for at least two reasons. One is that for large $n$ and $N$ the computations can grow as $\mathcal{O}(N n^3)$ and storage increases as $\mathcal{O}(n^2 + N)$. Due to the cubic complexity in $n$, there is a limit to the size of spatial data sets that can be easily analyzed with typical computing resources. For a laptop and R this limit is currently around several thousand for $n$ and tens of thousands for $N$, and prohibits the flexible and interactive analysis of larger spatial data sets. A second limitation is that inference for a spatial field is often a nonlinear function of the predictions and requires Monte Carlo sampling. This requires additional computation that is determined by both $n$ and $N$.

In this work, we propose a numerical algorithm for rapid spatial prediction onto a regular grid and for stationary covariance functions. This has a direct benefit of evaluating the Kriging prediction on a large grid to provide a smooth surface for visualization and for finding contours. Moreover, rapid prediction is also valuable for spatial inference when Monte Carlo computations are used to approximate the Kriging uncertainty. In either case, our goal is to provide accurate approximations that can give speedups on the order of a hundred or more. Our experience is that when this kind of efficiency is obtained it offers new ways of thinking about interactive data analysis and building statistical models.
Here, we leverage the fast numerical operation of convolution on a regular grid via the Fast Fourier Transform (FFT). This is done by approximating any covariance evaluation using just the covariance evaluated at grid points. Because our method is efficient, one can entertain a fine grid without much penalty for the extra computation time, and so it is easy to improve the accuracy of the approximations.

Conditional Simulation (CS) is a Monte Carlo-based algorithm that is well suited for quantifying Kriging prediction uncertainty.  The idea is that a Monte Carlo sample can be used to approximate the prediction variances or to find an empirical distribution for nonlinear functions of the predictions. A smaller number of Monte Carlo draws from the multivariate normal is often more efficient  to estimate the sampling distributions rather than direct analytical formulas. This strategy is akin to sampling from the posterior distribution in a Bayesian analysis rather than working directly with the posterior probability density function.
The key steps of CS are to (1) simulate an {\it unconditional} GP at the combined observation and prediction locations and (2) compute the Kriging prediction based on the simulated observations according to (1). Although we do not address Bayesian computations directly, it should be noted that CS is also intrinsic to sampling the posterior of the spatial process given posterior samples of the other statistical parameters.

CS becomes difficult to implement for large problems.For large prediction grids, unconditional simulation  becomes computationally intensive if observation locations are not also on the prediction grid. Previous work by Bailey et al. proposed a local simulation algorithm to provide fast, approximate unconditional fields for this first step \citep{Bailey2022} when the observations are at irregular locations. The success of the  local simulation algorithm, however, then identified the prediction step as a computational bottleneck. Reducing the time for this second step in CS is  the motivation for this work. It may come as a surprise to the reader that the spatial prediction step, essentially a multiplication of a (large) covariance matrix by a vector, is that important and merits so much attention. However, for the application of CS to large data sets it becomes the time-limiting computation. Without additional speedup interactive analysis is limited.  We term our new algorithm {\it rapid prediction} and report the accuracy and speedup for a range of spatial covariance models within the Mat\'{e}rn family. 
In most cases we achieve a factor of 100 or more speedup with acceptable accuracy. Moreover, we relate our local approximation idea to error bounds in approximation theory and so provide some theoretical understanding as to why our approach is numerically accurate. Previous work by \cite{Stein2011} has studied the locality of Kriging from a statistical standpoint, and using kernel approximations based on some interpolation theory is a complement to Stein's ideas. We also show these results hold for a practical, moderate-sized climate data set, mean summer rainfall data of the US. 

\section{Rapid Prediction for Kriging}\label{sec:RapidPredKriging}
Based on the spatial model (Eq. \ref{eq:ObservationalModel}), the standard Kriging prediction is derived by finding the best linear unbiased estimator for $f$. We start with inclusion of the linear part of the model, however, the numerical algorithms will focus on the spatial component because it has dimension on the order of the observations and prediction of the linear part is already efficient. Also we assume the covariance function and $\tau^2$ are known but discuss this issue in the last section. Throughout, for clarity, we focus on the spatial prediction for a two-dimensional domain. The extension to higher dimensions is straightforward although the computational speedup may not be as dramatic.

Let $k(\cdot, \cdot)$ be the covariance function for $g$ such that $k(\mathbf{s}, \mathbf{s}') ={\rm Cov} (g(\mathbf{s}), g(\mathbf{s}')) = \mathbb{E}(g(\mathbf{s}) g(\mathbf{s}'))$ for some arbitrary spatial locations $\mathbf{s}$ and $\mathbf{s}'$. Define the covariance matrix for the observation locations as $K_{i, l} = k(\mathbf{s}^O_i, \mathbf{s}^O_l)$ for $ i, l = 1, \dots, n$ and let $\mathsf{M} = K + \tau^2I_n$ where $I_n$ is the identity matrix with dimension $n$. The estimates for $\beta$ are found by generalized least squares (GLS) 
\begin{equation}
\label{eq: findbeta}
\hat{\beta} = (X^T \mathsf{M}^{-1} X)^{-1} X^T \mathsf{M}^{-1}\mathbf{z}
\end{equation}
where the $i$-th row of $X$ is $X_i$ and $\mathbf{z} = \{ z_i \}$ for $i = 1, \dots, n$. The coefficient vector is
\begin{equation}
\label{eq: findc}
\mathbf{c} = \mathsf{M}^{-1}( \mathbf{z} - X\hat{\beta})
\end{equation}
and the Kriging prediction at an arbitrary location $\mathbf{s}_j$ is given as
\begin{equation}
\label{eq:predictionh}
\hat{g}(\mathbf{s}_j) = \sum_{i=1}^n k( \mathbf{s}_j, \mathbf{s}^O_i) \mathbf{c}_i
\end{equation}
The complete prediction is
\begin{equation}
\label{eq:predictionf}
\hat{f}_j =  X_j\hat{\beta} + \hat{g}(\mathbf{s}_j).
\end{equation}

Although both (\ref{eq: findbeta}) and (\ref{eq: findc}) are given in terms of the inverse of $\mathsf{M}$, both of these computations can be made efficient. $\mathsf{M}$ is positive definite and one can evaluate $\mathsf{M}^{-1}$ via a Cholesky factorization and solving two triangular linear systems. Moreover, when $n$ is too large for an exact evaluation or storage of the matrix, one can use an approximate solution based on an iterative solution of large linear (and positive definite) systems. Thus we are led to consider the prediction step (\ref{eq:predictionh}). Without exploiting any additional structure or approximations this seemingly simple operation can dominate the Kriging prediction computation. Note that this operation needs to be done as many times as the number of CS samples, and so becomes a hurdle for computational speed. 
Next, we will discuss how we modified (\ref{eq:predictionh}) to speed up prediction.

\subsection{Rapid Prediction Methodology}\label{sec:Methodology}
To add structure to this problem we will assume that prediction is to a regular grid.
To streamline notation, however, we keep a single index reference to the prediction locations as $\{\mathbf{s}_j^G \}$ for $j = 1, \dots, N$. We also assume that $k(\cdot, \cdot)$ is stationary, so there is a function $\phi$ such that
\[ k(\mathbf{s},\mathbf{s}') = \phi(  \mathbf{s} - \mathbf{s}'). \]
This setup is typical of many spatial analyses and we suggest extensions to a nonstationary covariance
in the discussion. Our main idea is to approximate the function $k( \mathbf{s}^G_j,\mathbf{s}_i^O)$ as a linear combination of the kernel only evaluated at the grid points.
\begin{equation}\label{eq:ApproxCovariance}
	k\left( \mathbf{s}_j^G, \mathbf{s}_i^O\right)
	 \approx \sum_{q =1}^N  k\left( \mathbf{s}_j^G, \mathbf{s}_q^G\right) A_{i,q}
\end{equation}
Or more generally, for an arbitrary location:
\begin{equation}\label{eq:ApproxCovarianceGeneral}
	k\left( \cdot , \mathbf{s}_i^O\right) \approx \sum_{q = 1}^N  k ( \cdot , \mathbf{s}_q^G ) A_{i,q}
\end{equation}
Here, $A$ is a sparse matrix of interpolation weights that is described below. 

Let $\hat{g}_{approx} (\mathbf{s}_j^G)$ denote the resulting prediction substituting the approximation from (\ref{eq:ApproxCovariance}) into (\ref{eq:predictionh}).
\begin{equation}
\label{eq:ApproxPrediction}
	\hat{g}_{approx} (\mathbf{s}_j^G) =   \sum_{i = 1}^n  \left( \sum_{q =1}^N  k( \mathbf{s}_j^G, \mathbf{s}_q^G ) A_{i,q} \right)  \mathbf{c}_i
\end{equation}
Now setting $ \mathbf{c}^*_q = \sum_{i = 1}^n A_{i,q}\mathbf{c}_i $, interchanging the sums, and using the stationary covariance form for $k$, we get the following
\begin{equation}
\label{eq:ApproxPrediction2}
\hat{g}_{approx} (\mathbf{s}_j^G) =  \sum_{q =1}^N  k( \mathbf{s}_j^G, \mathbf{s}_q^G )  \left( \sum_{i = 1}^n A_{i,q}\mathbf{c}_i \right)
=  \sum_{q =1}^N  k( \mathbf{s}_j^G, \mathbf{s}_q^G ) \mathbf{c}^*_q =
\sum_{q =1}^N  \phi( \mathbf{s}_j^G - \mathbf{s}_q^G ) \mathbf{c}^*_q
\end{equation}
At this point it is not clear how this reformulation yields any computational savings! We exploit two features of this expression for rapid computation. First the approximation at (\ref{eq:ApproxCovariance}) is based on interpolation of $k(\cdot,\mathbf{s}^O_i)$ using only a small number, say $M$, of nearest neighbor grid points. Thus, finding the nonzero values of $A$ involves solving a small linear system and has $\mathcal{O}(M^3)$ complexity. Note that $A$ only needs to be found once for multiple $\mathbf{c}$ and so is well suited for CS. Subsequently $\mathbf{c}^*$ is found through the efficient sparse multiplication, $A ^T \mathbf{c}$, which has $\mathcal{O}(nM)$ complexity. The stationarity assumption for $k$ is used to convert the dense matrix multiplication in (\ref{eq:ApproxPrediction}) to a discrete convolution of $\phi$ with $\mathbf{c}^*$ in (\ref{eq:ApproxPrediction2}). This convolution over a regular grid is easily done through the FFT and is $\mathcal{O}(N \log N)$.
Thus we  convert the direct multiplication in (\ref{eq:predictionh}) to one that uses only sparse multiplication and the FFT and  this switch results in total complexity of $\mathcal{O}(N \log N + Mn + M^3)$ instead of exact prediction of $\mathcal{O}(Nn^3)$. The following subsection goes into details of this algorithm.

\subsection{Rapid prediction algorithm}\label{sec:Algorithm}

\begin{figure}
\vspace{-0.1in}
\centering{
\includegraphics[width=\textwidth]{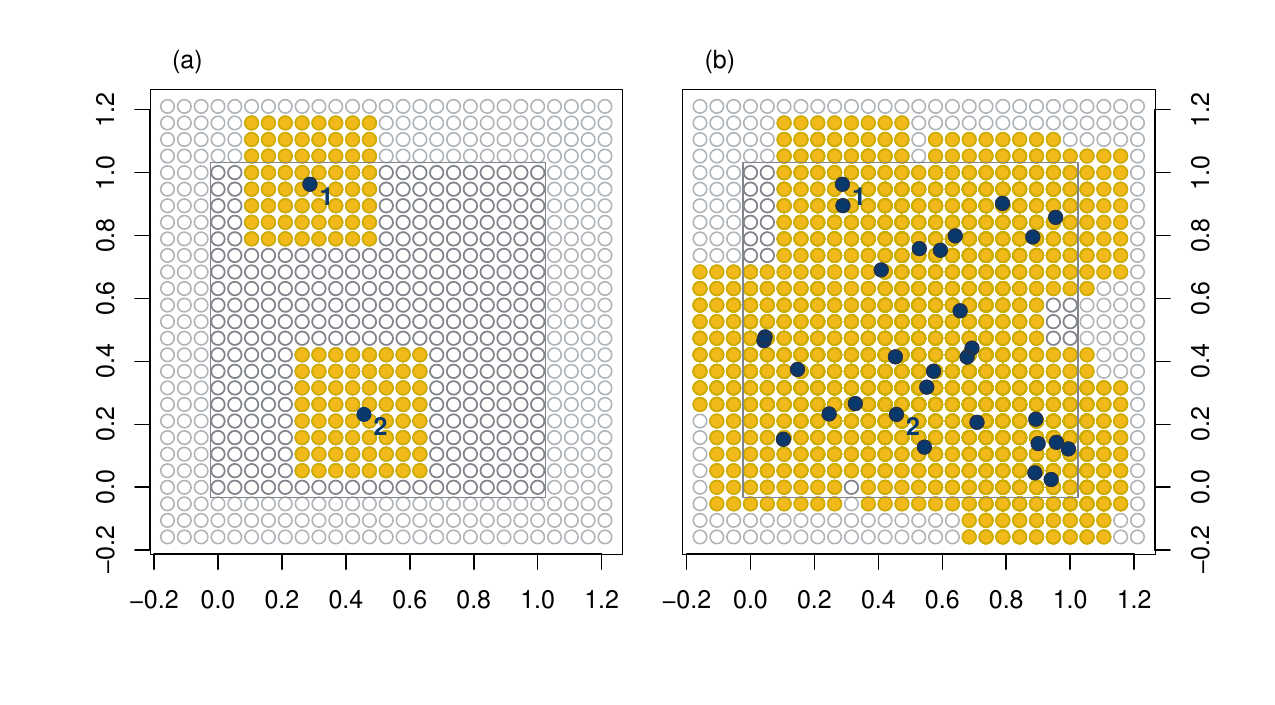}
}
\vspace{-0.6in}
\caption{\label{fig:VisualOfNNGridForObs} This figure illustrates the local approximation of the 30-observation example. White points within the rectangular outline are the prediction grid of interest and the remaining grid points are padding so that all locations have a full set of nearest neighbors. The reshaped $\mathbf{c}^*$ matrix would take the dimension of these combined gridded points. (a) highlights the two observations as dark blue points, and their respective $64$ nearest neighbors as gold points. The first row of sparse matrix $A$ would only contain nonzero values at the columns indexed by the yellow grid points around observation labeled ``1''. (b) shows all 30 observations at their off-grid locations $\{ \mathbf{s}_i^O\}$ in dark blue points and their collective nearest neighbors $\mathcal{N}$ in gold.}
\end{figure}

Similar to the algorithm for local Kriging, the computation is based around choosing a neighborhood of grid points around each observation location \citep{Stein2011}. Although this neighborhood can be chosen in several ways it is useful to focus on the one used in two dimensions and for the results in this work. Let $L$ be an integer that is the order of nearest neighbors around a location and so denotes a $2L \times 2L$ square of grid points. Note that this will comprise $M=(2L)^2$ points and demarcate the corners of $(2L-1)^2$ ``boxes'' with one box at its center. Finally, let $\mathcal{N}(i)$ be the subset of grid point indices for this $L$-th order nearest neighborhood where $\mathbf{s}^O_i$ is contained in the central box (see Figure \ref{fig:VisualOfNNGridForObs}).

We define the nonzero entries, $A_{i,q}$, for $A$ as the coefficients that solve the interpolation condition:
\begin{equation}
\label{eq:weights}
 k(  \mathbf{s}_j ^G, \mathbf{s}_i ^O) =  \sum_{q \in \mathcal{N}(i)} k( \mathbf{s}_j^G, \mathbf{s}_q^G) A_{i,q}
 \end{equation}
for $j \in \mathcal{N}(i)$. $A_{i,q}$ is set to zero for $q \notin \mathcal{N}(i)$, hence inducing the sparsity of $A$. The interpolation condition is easiest to explain in terms of covariances of a GP.
 Let $\mathbf{k}_i$ be the vector of covariances between $\mathbf{s}_i^O$ and $\mathbf{s}_j^G$ where $j \in \mathcal{N}(i)$ --- the covariance between the off-grid observation location and its nearest neighbors. Let $K_i$ be the covariance matrix among the nearest neighbors of $\mathbf{s}_i^O$. Then we have
\[
 K_i A_{i} = \mathbf{k}_i
 \mbox{ or }
 A_{i} = K_i^{-1}  \mathbf{k}_i^T
 \]
where $A_i$ is the $i^{th}$ row of $A$ restricted to its nonzero values (i.e. the vector of $A_{i,q}$ values where $q \in  \mathcal{N}(i)$). When $k$ is stationary $K_i$ will be the same independent of $i$ and so can be explicitly inverted once and used for finding all $n$ rows of $A$. In this case denote this matrix as $K_\mathcal{N}$, dropping the subscript.

Now, with the new prediction weights $\mathbf{c}^* = A^T \mathbf{c}$, we can rewrite (\ref{eq:ApproxPrediction2}) as a discrete convolution of the filter based on $\phi$ and applied to the ``image'' matrix $\mathbf{c}^*$. Using $\ast$ to denote convolution, $\Phi$ the discrete filter, then
\begin{equation}
\label{eq:Convolution}
\hat{g}_{approx} (\{ \mathbf{s}_j^G\}) = \Phi \ast \mathbf{c}^*
\end{equation}
computes the rapid approximate prediction on the set of all prediction grid locations of interest $\{ \mathbf{s}_j^G\}$ for $j = 1, \dots, N$ assuming observations are removed from the edges of the spatial domain.

Figure \ref{fig:VisualOfNNGridForObs} is a small, illustrative example setting up rapid approximate prediction for 30 observations onto a $20 \times 20$ grid of $[0, 1] \times [0, 1]$ and nearest neighbor size of $L = 4$. The grid has been extended (padded) beyond the spatial domain so that all observations have the full number of nearest neighbor grid points. In this example the padding is 3 extra points in the horizontal and 4 in the vertical, thus adding $N= 768 = 368 +400$. The local approximation for each covariance kernel centered on the observation $\mathbf{s}^O_i$ involves the $(2 \times 4)^2 = 64$ nearest neighbor grid points. The nearest neighbor covariance matrix $K_i \equiv K_\mathcal{N}$ is a $64 \times 64$ matrix and complete sparse weight matrix $A$ has dimensions $30 \times N $.

With this motivation, the rapid prediction algorithm is given below for the two-dimensional approximate prediction with $n$ observations, using $L$ nearest neighbors. We also assume the grid is large enough or has been extended so that all observations have $L$ nearest neighbors. Assume that the prediction grid has $m_1$ grid points in the first coordinate and $m_2$ in the second,
and both are odd integers and so by our definition $N= m_1 m_2 $. Finally assume $\mathcal{F}$ and $\mathcal{F}^{-1}$ denote the discrete two dimensional FFT and its inverse.
Below we outline the full algorithm for rapid prediction for a stationary covariance.

\begin{enumerate}
\item   Compute $K_\mathcal{N}^{-1}$
\item   Compute the discrete $N\times N$ filter $\Phi$ based on $\phi$
  and embed this array into an $(2m_1 -1)  \times (2m_2-1)$ circulant array, $\tilde{\Phi} $.
\item  for $i = 1, \dots, n$ \  \
  $\mathbf{k}_i = \{ k(\mathbf{s}^O_i, \mathbf{s}^G_q)  :   q \in \mathcal{N}(i) \}$
\item  for $i = 1, \dots, n$   \  \
  $A_i =  (K_\mathcal{N})^{-1}\mathbf{k}_i$
 \item  Sparse matrix multiply:  $\mathbf{c}^* = A^T \mathbf{c}$
 \item    Reshape $\mathbf{c}^*$ as a two-dimensional, $m_1 \times m_2$ array. Embed this result into the leading elements of  an $(2m_1-1)\times (2m_2-1)$ array, $\tilde{\mathbf{C}}$ with the remaining entries padded as zero.

 \item  $  \hat{g}_{approx}^G = {\mathcal{F}}^{-1}\left( {\mathcal{F}}(\tilde{\Phi})  {\mathcal{F}}(\tilde{\mathbf{C}})  \right)$

 \end{enumerate}
 
 Some remarks are in order for these steps. First the reader will note that the matrix $(K_\mathcal{N})^{-1}$ only needs to be computed once for looping over $i$ if the ordering of nearest neighbors is the {\it same} for every $\mathcal{N}(i)$, hence it makes sense to use the single symbol $K_\mathcal{N}$ for all neighborhoods. Furthermore, $\Phi$, $A$ and $\mathcal{F} (\tilde{\Phi})$ also only need to be computed once for multiple $\mathbf{c}$. For clarity we have omitted some notational rigor in this algorithm concerning the indexing. In Step 5 the sparse matrix, $A$, has been filled row by row in Step 4 with the nonzero columns that are the nearest neighbors to $i$. Formally the multiplication in Step 5 assumes the zeroes have been identified for the entries of $A$ so this makes sense as a single sparse matrix multiplication. The final step returns an $(2m_1 -1)  \times (2m_2-1)$ array due to the circulant extension although the actual predictions are the first $m_1 \times m_2$ block of this array.

The standard circulant construction in Step 2 of $\tilde{\Phi}$ using reflections of the original matrix may seem awkward and possibly unnecessary. However, this ensures that the convolution at the edges is handled correctly. In most cases the FFT affords enough efficiency so that increases of the array size by a factor of two is not an issue. The construction of $\tilde{\Phi}$ becomes tedious for higher dimensions and an alternative computation can avoid the reflections and indexing but requires an additional FFT to set up $\tilde{\Phi}$. This implementation is in the fields function {\tt circulantEmbeddingSetup} along with an adjustment for further speedup in the array size to ensure the arrays have dimensions that are highly composite -- only factors of 2 and 3.

\section{Approximation Error}\label{sec:ApproximationError}
Recall the difference between rapid prediction and the exact method at any arbitrary location $\mathbf{s}$ is
\begin{equation}
\label{error1}
	\hat{g}_{approx}(\mathbf{s}) - \hat{g}(\mathbf{s}) = \sum_{i = 1}^n k_{approx}(\mathbf{s}, \mathbf{s}_i^O) \mathbf{c}_i - \sum_{i = 1}^n k(\mathbf{s}, \mathbf{s}_i^O) \mathbf{c}_i = \sum_{i=1}^n \left( k_{approx}(\mathbf{s}, \mathbf{s}_i^O) -k(\mathbf{s}, \mathbf{s}_i^O) \right) \mathbf{c}_i
\end{equation}
with $k_{approx}$ following (\ref{eq:ApproxCovarianceGeneral}) and the first sum being refactored to exploit the FFT. The expected value
of this difference
is zero based on the usual spatial model where $\mathbb{E} \left[ \mathbf{c}_i \right] = 0$. Thus we are led to consider the  absolute value of the error: $| \hat{g}_{approx}(\mathbf{s}) - \hat{g}(\mathbf{s})|$. 

In this section we analyze the error numerically with a Monte Carlo study to determine which factors of this problem have the most influence on accuracy. These results motivate considering analytical expressions for interpolation error bounds.

The isotropic Mat\'{e}rn covariance family is used throughout this section as the baseline covariance model for quantifying approximation error.
This common family of covariance functions is defined by
\begin{equation}\label{eq:Matern1}
k(\mathbf{s}, \mathbf{s}') =  \sigma^2 \phi (  \alpha \Vert \mathbf{s} - \mathbf{s}' \Vert )
\mbox{ and }
\phi(d) = \frac{ (d)^\nu \mathcal{K}_{\nu} ( d ) } {2^{\nu - 1} \Gamma(\nu)}
		\end{equation}
where $\sigma^2$ is the process variance, $\alpha$ is a scale (range) parameter, $\nu$ is the smoothness, $\mathcal{K}_\nu$ is an order $\nu$ Bessel function of the second kind, and $\Gamma$ is the gamma function \citep{Lindgren2011}.

\subsection{Empirical Approximation Errors }\label{sec:EmpiricalApprox}
We explore the expected prediction error in (\ref{error1}) through a factorial design via Monte Carlo simulations.
We consider the following factors that may influence accuracy: number of nearest neighbors, the spacing of grid points on a $[0, 1] \times [0,1]$ domain, number of observations, and kernel properties such as Mat\'{e}rn smoothness and range parameters. These factors were studied through a $3^6$ factorial design with $729$ individual cases.
The factors with their  levels are listed below:
\begin{itemize}
	\item Number of observations $n : 200, 500, 1500$.
	\item Mat\'{e}rn range parameters determined as a correlation of 0.7 at distances: $0.2, 0.4, 0.8$.
	\item Mat\'{e}rn smoothness $\nu: 0.5, 1, 1.5$.
	\item Nugget variance $\tau^2: 0.01, 0.1, 0.5$.
	\item Nearest neighbor order $L: 2, 4, 8$.
	\item Prediction grid sizes:$100 \times 100, 350 \times 350, 500 \times 500$.
\end{itemize}
The number of simulated data observations and nugget variance were chosen to be close to the rainfall example in Section \ref{sec:RainfallExample}. The Mat\'{e}rn smoothness values were chosen to bracket the rainfall example $\nu = 1$, and also correspond to connections to kernel approximation theory in Section \ref{sec:KernelTheory}; correlation distances were chosen to correspond to the rainfall example and to account for a variety of situations that might occur in similar applications. The nearest neighbor sizes were chosen to include our preferred  value of $L = 4$ and the example in Section \ref{sec:KernelTheory} where $L = 2$. Finally the prediction grid sizes  correspond to the timing examples we will discuss in Section \ref{sec:Timing}.

\begin{figure}
\vspace{-0.5in}
\centering{
\includegraphics[width=0.5\textwidth]{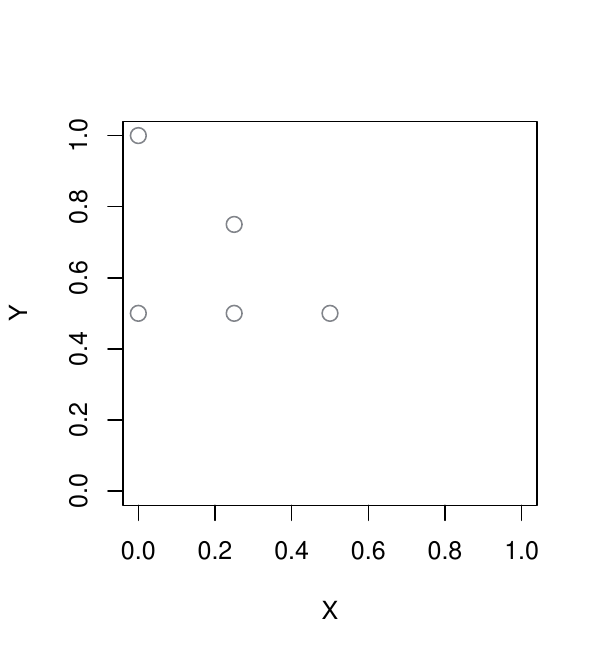}
}
\vspace{-0.25in}
\caption{\label{fig:MC_EvalPoints} Prediction locations from numerical study.}
\end{figure}%

The accuracy of the approximation will depend on the location of observation data and where the prediction is evaluated. To control for these two features for each sample size we choose $50$ sets of observation locations uniformly distributed from the square $[0, 1] \times [0,1]$. The error is averaged over these 50 cases.
 We choose five prediction points for evaluation, varying from the center to the corner and edges of the spatial domain (see  Figure~\ref{fig:MC_EvalPoints}). This allows us to study how the approximation might degrade at edge or corner points. For each of the prediction locations and each combination of levels we find average absolute error for the $50$ simulated data sets. The $\log$ of this average is used in the ANOVA summary.

\begin{figure}
\centering{
\includegraphics[width=\textwidth]{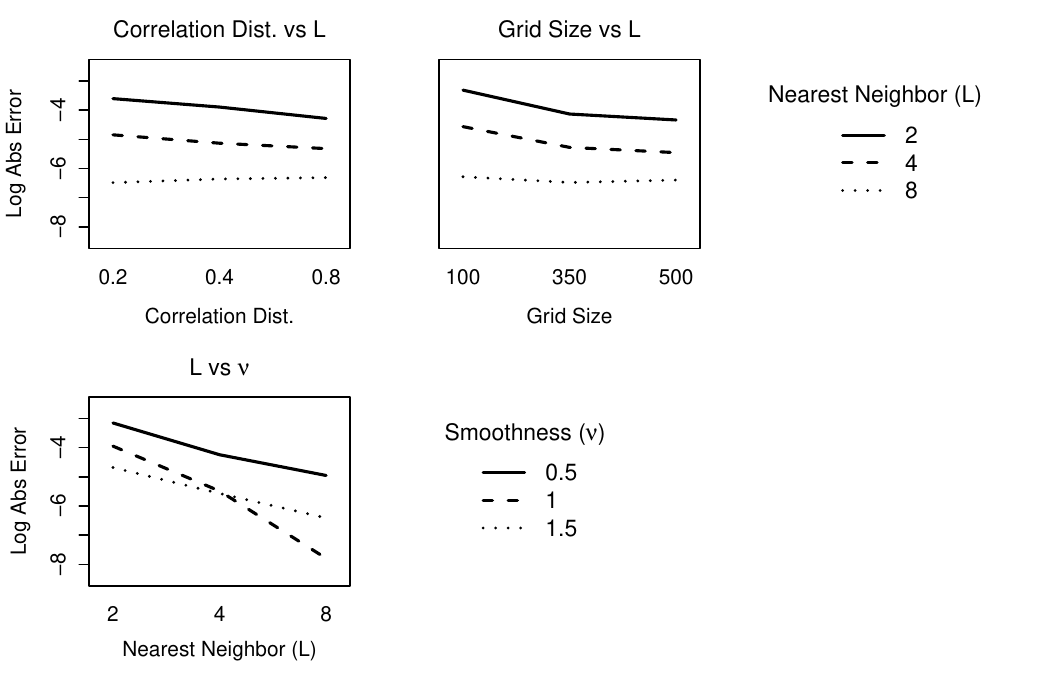}
}
\vspace{-0.4in}
\caption{\label{fig:InteractionPlot} Interaction plots of the factorial design study showing interactions that were identified as being potentially important from the full model.}
\end{figure}%

An analysis of variance (ANOVA) of the complete results is reported in Appendix~\ref{appendix:a}, Table~\ref{tab:fullANOVA}. It indicates that the vast majority of variability in $\log$ expected absolute error can be explained by smoothness, nearest neighbor size, the interaction between them, the number of prediction grid points, and the nugget variance. The interaction plot (Figure~\ref{fig:InteractionPlot}) gives the reader an idea of the size of the interactions with the strongest interaction patterns between smoothness and nearest neighbor size. Hence, to simplify the summary we just include the most influential factors in a smaller model. Surprisingly, the results show that the rapid approximate prediction method accuracy does not depend on the prediction locations. We believe this is due to the algorithm adding padding to edge observations that effectively extends the prediction domain, hence edge effects do not apply. Note that this is different than the spatial prediction error itself which does  increase at corners and edges relative to the center of the domain. 

\begin{longtable}[]{@{}llll@{}}
\caption{ANOVA result on the reduced model of log absolute error based on smoothness ($\nu$), nearest neighbor order ($L$), grid size, nugget variance ($\tau^2$), and the interaction between $\nu$ and $L$ of the MC experiment.}\label{tab:reducedANOVA}\tabularnewline
\toprule\noalign{}
Factor & Degrees of Freedom & Mean Square \\
\midrule\noalign{}
\endfirsthead
\toprule\noalign{}
Factor & Degrees of Freedom & Mean Square \\
\midrule\noalign{}
\endhead
\bottomrule\noalign{}
\endlastfoot
Mat\'{e}rn Smoothness $(\nu)$              & 2 & 192.22 \\
Nearest Neighbor $(L)$                    & 2 & 367.59 \\
Grid Size                                 & 2 &  32.38 \\
Nugget Variance $(\tau^2)$                & 2 &  30.74 \\
Smoothness $\times$ Nearest Neighbor      & 4 &  31.63 \\
Residuals                                & 716 &   0.20 \\
\end{longtable}

From Table~\ref{tab:reducedANOVA}, we see that the reduced model is sufficient to explain the variability in the MC experiment with the mean square of residuals being less than 1\% of the mean square variation for each factor. The detailed estimate results (Appendix~\ref{appendix:a}, Table~\ref{tab:reducedLM}) further show that in the worst case of $\nu = 0.5$, $\tau^2 = 0.01$, and $L = 2$, the approximation is accurate to at least 2 decimal places with average log absolute error being less than $10^{-2}$. With everything else held constant, an increase of smoothness from $\nu = 0.5$ to $\nu = 1.5$ results in more than 1 decimal place of improvement in accuracy. Similarly, increasing the nearest neighbor order from $L = 2$ to $L = 4$ also improves accuracy by 1 decimal place on average. Increasing grid size and nugget variance also improve accuracy but less dramatically than increasing nearest neighbor order or smoothness. Interestingly, the most drastic improvement from the experiment is in the interaction of $\nu = 1$ and $L = 8$, where slightly increased smoothness and a large increase in nearest neighbor order results in 2 decimal places of improvement in accuracy.

\subsection{Connection to Kernel Approximation Theory}\label{sec:KernelTheory}
The empirical results in the previous section motivate closer study of how the approximation improves with smoothness and grid spacing.
 For that, we modify (\ref{error1}) and get the elementary inequality
\[
 	| \hat{g}_{approx}(\mathbf{s}) - \hat{g}(\mathbf{s})|  \le
	  \left\{  \sup_{i}   | k_{approx}(\mathbf{s}, \mathbf{s}_i^O) -k(\mathbf{s}, \mathbf{s}_i^O) |  \right\} \sum_{i=1}^n | \mathbf{c}_i |
\]
and so
\begin{equation}
 	\mathbb{E} \left[ | \hat{g}_{approx}(\mathbf{s}) - \hat{g}(\mathbf{s}) | \right] \leq
\left\{  \sup_{i}   | k_{approx}(\mathbf{s}, \mathbf{s}_i^O) -k(\mathbf{s}, \mathbf{s}_i^O) |  \right\}
 \mathbb{E} \left[ \Vert \mathbf{c} \Vert_1 \right].
	  \label{eq:absError}
\end{equation}

The maximum absolute error between the rapid prediction algorithm and the exact method changes based on the different variables involved. It is not surprising that the largest error occurs when an observation location is {\it centered} in the central grid box formed by the nearest neighbor grid (see Figure~\ref{fig:DiagramFillDistance} below) and empirically we confirmed this intuition. So we restrict this analysis to this case for $\mathbf{s}_i^O$ centered in the approximation grid box. Given this choice for the hypothetical observation location we then find the maximum error for $\mathbf{s}$ in the spatial domain. Finally, due to our assumptions of isotropy and the padding of the prediction grid we also focus on the error when the observation location is in the center of the prediction domain. With these reductions we study the error, $\Lambda(\Omega, \mathbf{s}^*)$.
\begin{equation}
 \Lambda(\Omega, \mathbf{s}^*) = \sup_{\mathbf{s} \in \Omega}  | k_{approx}(\mathbf{s}, \mathbf{s}^*) - k(\mathbf{s}, \mathbf{s}^*) |
\end{equation}
where $\Omega$ is the prediction domain and $\mathbf{s}^*$ the location in the center of the central grid box.

Standard approaches for interpolating curves and surfaces can be formulated as minimization (variational) problems over norms on the interpolating function. In this section we switch to slightly different notation to be consistent with \cite{Fasshauer2006, Fasshauer2007}. In particular, $h(\mathbf{x}) $ is now a smooth function that is to be estimated/interpolated from observations and the locations are now $\mathbf{x}$. Below we sketch this theory leading to a bound on $\Lambda(\Omega, \mathbf{x})$ that depends as a polynomial in the grid spacing and with degree controlled by the smoothness of the
 covariance (kernel).

Let $\mathcal{H}$, be a Hilbert space of functions $h: \mathbb{R}^d \rightarrow \mathbb{R}$ with reproducing kernel (RK) $\Phi: \mathbb{R}^d \times \mathbb{R}^d \rightarrow \mathbb{R}$, norm $\| \cdot \| $ and inner product $\langle \cdot, \cdot \rangle$ (see Appendix \ref{appendix:b} for details). The linear combinations of $\{\Phi(\cdot, \mathbf{x}) \}$ form a dense set of $ \mathcal{H}$ and we have the reproducing property
$$ \langle h , \Phi(\cdot, \mathbf{x}) \rangle =  h(\mathbf{x})$$
 for all $ h \in \mathcal{H}$. Taken together we refer to this combination of $ \langle . , . \rangle $ and $\Phi$ as a {\it reproducing kernel Hilbert space} (RKHS).
 
 Let $\{ (\mathbf{x}_i , \mathbf{y}_i )\}_{i=1}^n$ be values for some unknown function $h$ where $h(\mathbf{x}_i) = \mathbf{y}_i$. The goal is to find a function $ \hat{h} \in \mathcal{H}$ that approximates $h$ based on these discrete values. The estimate we consider is the function in $\mathcal{H}$ that minimizes $\| h \| $ over all functions $ h \in \mathcal{H}$
 and that satisfies the constraint $ h(\mathbf{x}_i) = \mathbf{y}_i$ for $ i = 1, \dots, n$. A solution exists, is unique, and has a finite-dimensional form. Moreover, based on the correspondence between splines and Kriging \citep{kimeldorf_1970}, $\hat{h}$ will also be the Kriging solution under the assumption of $\tau=0$ and $\Phi = k$.
 
 In a RKHS, the interpolation error $ | h(\mathbf{x}) - \hat{h}(\mathbf{x}) |$ has a simple and elegant error bound
\[  | h(\mathbf{x}) - \hat{h}(\mathbf{x}) |   \le \mathbf{P}_\Phi (\mathbf{x})  \| h \|_{\cal H}     \]
with $\mathbf{P}_\Phi (\mathbf{x})$ first identified by \cite{Schaback1993} as the \textit{kriging function} and later termed the {\it power function}. The benefit of this bound is that we can bound the error by two components: $ \| h \|_{\cal H}$ which is dependent on $\Phi$ and $h$ but {\it not} on data locations, and the power function that depends on $\Phi$ and the locations of the discrete values, but not on $h$.
Moreover a bound on the power function can be found using the quantity the {\it fill distance, $\delta$}.
Let
\begin{equation}
\label{minDist}
d( \mathbf{x}, { \mathbf{x}_i} ) =  \min_{i} \| \mathbf{x} - \mathbf{x}_i \|
\end{equation}
Then fill distance is
\begin{equation}
\delta =  \sup_{\mathbf{x} \in \Omega} d( \mathbf{x}, \{ \mathbf{x}_i\} )
\end{equation}
Statisticians may recognize this as the ``minimax'' criterion for generating a space-filling design. (To find the best design one would vary $\{\mathbf{s}_i\}$ to minimize $\delta$ as this is a measure of how well $\{\mathbf{x}_i\}$ cover, or fill, $\Omega$.)

\begin{figure}
\centering{
\includegraphics[width=0.6\textwidth]{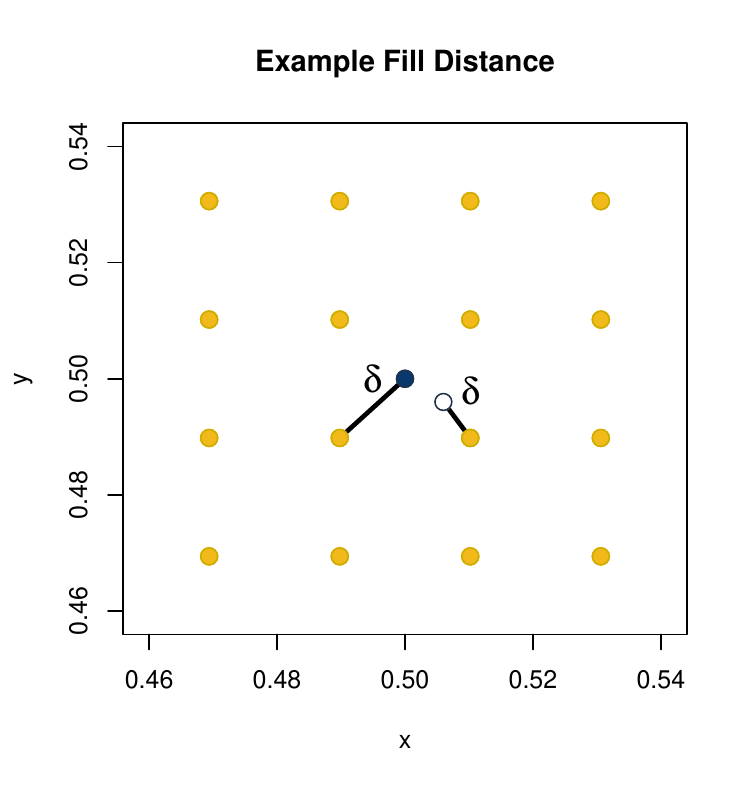}
}
\vspace{-0.4in}
\caption{\label{fig:DiagramFillDistance} An illustration of the minimum distance in (\ref{minDist}) with the interpolation grid points (yellow) and two hypothetical observations and labeled distances being the minimum for the observation to its nearest grid point. Visually it is apparent the maximum ($\delta$) must be for the centered location.}
\end{figure}%

For $\delta$ sufficiently small we have the simplified bound (more details in Appendix~\ref{appendix:b})
\begin{equation*}
| h(\mathbf{x}) - \hat{h}(\mathbf{x}) |   \le C(\mathbf{x}) \delta^\kappa  \| h \|_{\cal H}.
\end{equation*}
Now taking the supremum over $\mathbf{x}$ we get
\begin{equation}\label{eq:supKernelBound}
\sup_{\mathbf{x} \in \Omega} \left| h(\mathbf{x}) - \hat{h}(\mathbf{x}) \right| \le C \delta^\kappa
\end{equation}
where $\delta$ is the fill distance, $\kappa$ the approximation order of the kernel,
and $C$ is a constant depending on $\{ \mathbf{x}_i\} $ and $h$. To map this to the Matern family of kernels in 2 dimensions $\kappa = \nu- 1/2$ \citep{Fasshauer2007}. Finally to apply this to the rapid prediction error in
$\Lambda(\Omega, \mathbf{s}^*)$ we identify $ h \equiv k( . , \mathbf{s}^*)$ and $\hat{h} \equiv k_{approx}(. , \mathbf{s}^*)$.

In summary, the analytical bound for approximating the kernel is a polynomial function of $\delta$ and should exhibit errors that on a log-log scale are linear with respect to $\log( \delta)$.
As is common in approximation theory it is convenient to consider the relationship with respect to $1/\delta$ and so we expect a slope of $-\kappa$ on a log-log scale. Note that the grid spacing is proportional to $\delta$ and so $1/\delta$ is proportional to the number of grid points in the domain.
  
\begin{figure}
\centering{
\includegraphics[width=0.95\textwidth]{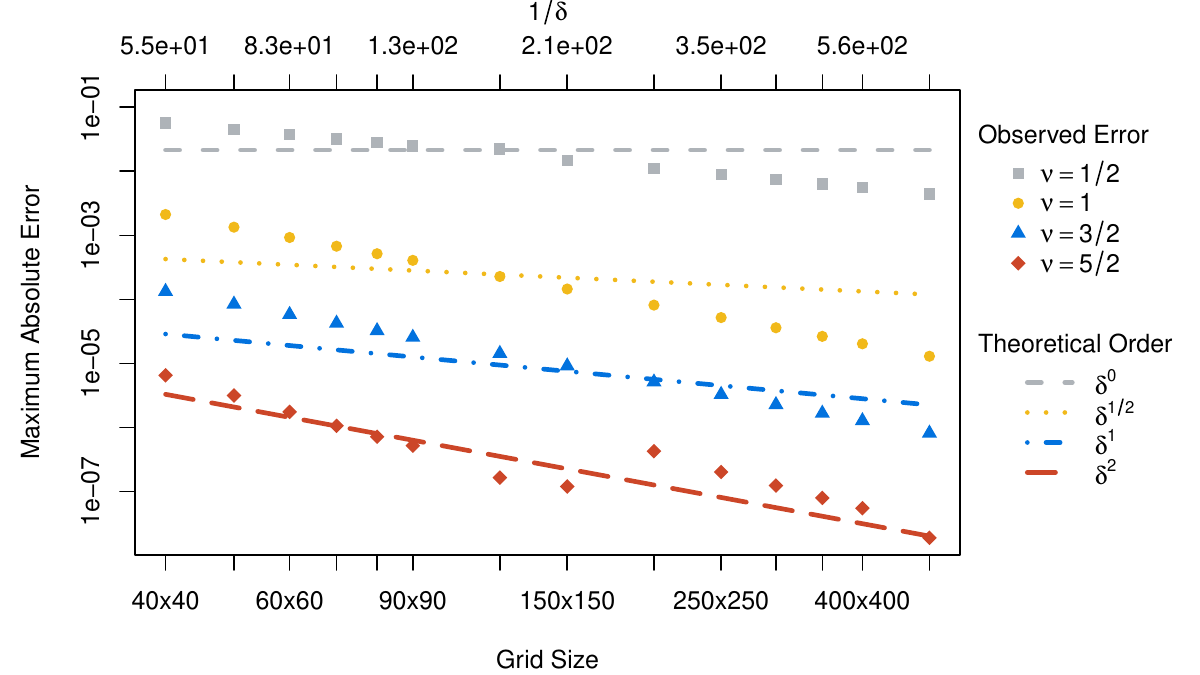}
}
\vspace{-0.17in}
\caption{\label{fig:FillDistanceMaxError} Maximum absolute error for approximating $k(\cdot, \mathbf{s}^*)$ using $k_{approx}(\cdot, \mathbf{s}^*)$ over inverse fill distance over different Mat\'{e}rn smoothnesses of $\nu = 1/2, 1, 3/2$ and $5/2$, a fixed range of $\alpha = 0.25$ and nearest neighbor order of $L = 2$. Theoretical expected approximation orders are shown in lines with color corresponding to each $\nu$.}
\end{figure}%
  
Figure~\ref{fig:FillDistanceMaxError} is a log10-log10 plot of this error as a function of $1/\delta$
 for different smoothness values, and the scale parameter fixed at $0.25$.
 Superimposed are the error bounds from the analytical results. The empirical and theoretical slopes are reported in Table~\ref{tab:orderConvergence}. The slopes for the numerical results are slightly higher
 than theoretical ones but the match improves as the smoothness increases. However, there is also some
 numerical instability for grid sizes larger than $200 \times 200$ for $\nu = 5/2$. In general, the linearity of the log errors is striking and a confirmation that the theoretical bound is useful for insight into this algorithm.

\begin{longtable}[]{@{}lllll@{}}
\caption{Mat\'{e}rn smoothnesses, their expected order in 2 dimensions and actual slope from the example found by least squares. $^{*}$ The least squares for the $\nu = 5/2$ case was a poor fit due to numerically instability for  fill distances corresponding to the $200 \times 200$ and finer prediction grid cases; slope for $140 \times 140$ and coarser grid is 2.99.}\label{tab:orderConvergence}\tabularnewline
\toprule\noalign{}
\textbf{Smoothness $\nu$} &  \textbf{Kernel Continuity} & \textbf{Expected $\kappa$} & \textbf{Empirical $\kappa$} \\
\midrule\noalign{}
\endfirsthead
\toprule\noalign{}
\textbf{Smoothness $\nu$} &  \textbf{Kernel Continuity} & \textbf{Expected $\kappa$} & \textbf{Empirical $\kappa$} \\
\midrule\noalign{}
\endhead
\bottomrule\noalign{}
\endlastfoot
1/2    &  $C^0$         & 0    &	1.00    \\ \hline
1	&  $C^1$		& 1/2 & 	1.99 	\\ \hline
3/2    & $C^2$         & 1     &	 2.00       \\ \hline
5/2	& $C^4$	& 2	&	1.84 (2.99)$^{*}$ 
\end{longtable}

The actual order of convergence tends to be slightly greater than theoretical $\kappa$ but this agreement improves as smoothness increases. This feature is well known in kernel approximation where the numerical error tends to be smaller than that predicted by theory.
 
 The spacing on the x-axis of Figure~\ref{fig:FillDistanceMaxError} has several interpretations and it is useful to describe these. We generated this figure assuming a scale parameter of $0.25$ and varied the grid spacing, $\delta$, in the nearest neighbor approximation. Here the domain is fixed at $[0, 1] \times [0, 1]$ and so the implied number of grid points is $1/\delta$. Varying the range parameter from $0.25$ is equivalent to adjusting the grid spacing. For example, the approximation error at $\delta$ and a range of $0.25$ is identical to the error for a scale parameter of $0.5$ and a spacing of $2\delta$ -- it is the same interpolation problem. In this way the accuracy reported in this figure is not limited to a single choice of range parameter. 

\section{Timing Results}\label{sec:Timing}
For timing we measure just the prediction method and also when it is implemented into the conditional simulation scheme. In both cases, we generate 200, 1600, and 6500 observations uniformly sampled from $[0,1] \times [0,1]$ with a Mat\'{e}rn covariance function with smoothness 1 and range 0.05. Prediction was done onto evenly spaced grids of sizes $60 \times 60$ to $500 \times 500$. Note that the choice of the smoothness and range parameter does not affect the timing and there is only a slight dependence on the observation locations based on overlap of nearest neighborhoods.

\subsection{Timing for Prediction Only}\label{sec:TimingPrediction}

\begin{figure}
\centering{
\includegraphics[width=\textwidth]{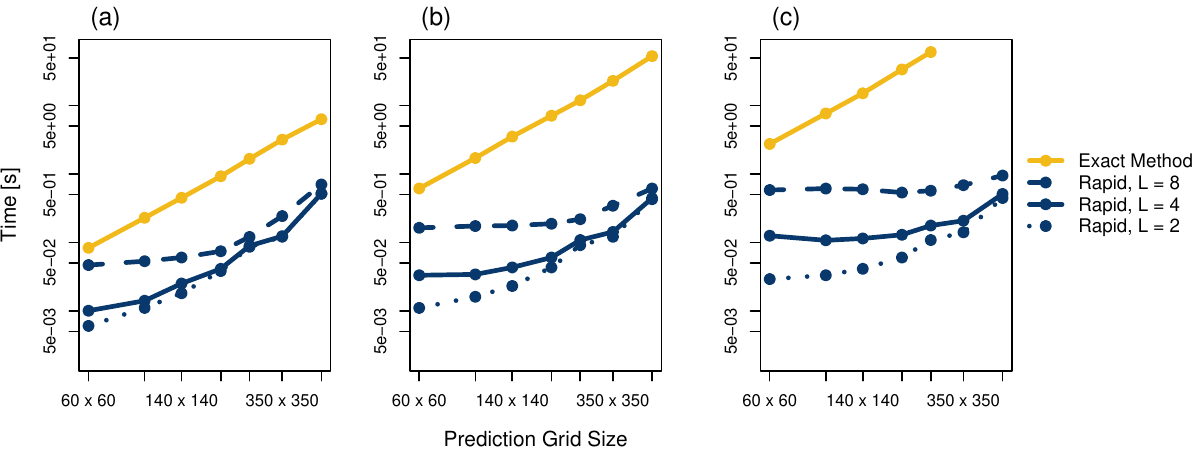}
}
\vspace{-0.3in}
\caption{\label{fig:SimPredictionTiming} Median time for prediction over up to $500 \times 500$ prediction grid points. Computations done on (a) presents timing results for 200 observations, (b) for 1500 observations, and (c) for 6500 observations.}
\end{figure}%

For the rapid prediction method we consider nearest neighbor orders of $L = 2, 4$, and $8$. All predictions use the implementations in the {\tt predictSurface} function from the \texttt{fields} R package and timings are reported for an Apple M2 Processor with a MacOS-specific BLAS ({\tt libRblas.vecLib.dylib}).

Figure \ref{fig:SimPredictionTiming} reports the timing results over these cases. Although the start-up cost of inverting the nearest neighbor grid can be high for a large number of nearest neighbors ($L$) for the rapid method, the efficiency of the convolution of order $\mathcal{O}(N \log N)$ with a large number of prediction grid points $N$ results in a reduced amount of total time. As expected with a larger number of observations, $n$, the speedup is dramatic -- being on the order of a factor of 100 or more.

\subsection{Timing in Conditional Simulation}\label{sec:TimingCS}
\begin{figure}
\centering{
\includegraphics[width=\textwidth]{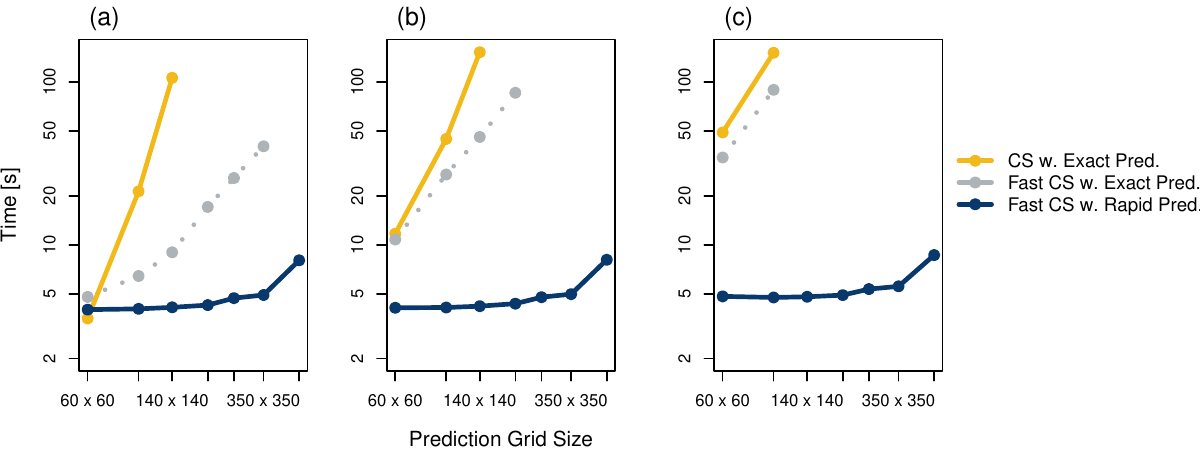}
}
\vspace{-0.3in}
\caption{\label{fig:SimCSTiming} Shows timing for a 10-member ensemble of conditional simulation using exact conditional simulation , the approximate  scheme using order 4 local simulation \citep{Bailey2022}, and further approximation conditional with order 4 local simulation and  rapid prediction method with nearest neighbor order $L = 4$. (a) presents timing results for 200 simulated observations, (b) for 1500 observations, and (c) for 6500 observations.}
\end{figure}%

The original inspiration for this project is addressing the bottleneck for a faster approximate conditional simulation scheme. The rapid approximation method has been incorporated into this algorithm. Here we give some timing comparisons for  three different variations of conditional simulation: exact simulation, the fast conditional simulation scheme \citep{Bailey2022} but using exact prediction, and the fast conditional simulation scheme with rapid prediction using nearest neighbor order, $L= 4$. All three implementations are in the function {\tt sim.spatialProcess} from the \texttt{fields} R package and 10 draws are sampled.  

Figure \ref{fig:SimCSTiming}, gives timing results across these methods, sample sizes and grid sizes. We see that there is no benefit in the smallest problem ($n=200, N= 60$); however, the speedup becomes dramatic as $N$ increases. As expected for large sample sizes the conditional simulation is dominated by the prediction step and so we achieve the speedup by inclusion of rapid prediction. Note that in the context of this computation the setup cost for the rapid prediction becomes negligible because it is only done once. Similar to the prediction-only results we have achieved orders-of-magnitude speedup. Moreover from Section 3 we know the method is not only fast but accurate.

\section{Annual Mean Rainfall Example}\label{sec:RainfallExample}

As a final illustration of the rapid prediction method we give an example from the \texttt{NorthAmericanRainfall2} dataset in the \texttt{fields} package. This dataset entails mean summer
(June, July, August) rainfall between 1971 - 2023 from  the Global Historical Climate Network (GHCN) version 4 \citep{fields} and has 4893 locations over North America. The goal is to quantify the difference in rainfall patterns between the Eastern and Western United States, exploring the historical dividing line of the 100-th
meridian.This line of longitude was informally characterized as the division between conventional agriculture and that requiring irrigation or other dryland methods. Thus we will only examine a smaller and more manageable subset of this dataset between longitude of $[-105.0, -92.0]$ and latitude of $[27.0, 55.0]$. This subset has $n = 1368$ observations. The contour at a specific level is a nonlinear function of the data and so this feature is an example of where conditional simulation is required to quantify its uncertainty.

We fit a spatial model with fixed effect being linear regression of longitude, latitude, and their interaction, with their regression coefficients found by GLS and a Mat\'{e}rn covariance function with smoothness fixed at $\nu = 1.5$. The resulting model has maximum likelihood estimates scale/range $\alpha = 1.21$, process variance $\sigma^2 = 2.43$, and nugget variance $\tau^2 = 0.47$. Using this model, we compare the rapid approximate prediction method to exact prediction and also timing against two alternative spatial models that enjoy efficient computation: Vecchia approximation \citep{GpGp} and LatticeKrig \citep{LatticeKrig}.

\subsection{Accuracy}\label{sec:RainfallAccuracy}

We compare prediction accuracy by applying the exact method and rapid method with the nearest neighbor order $L = 4$ using the same fitted spatial model onto an evenly spaced $256 \times 512$ grid over the data domain.

\begin{figure}
\centering{
\includegraphics[width=\textwidth]{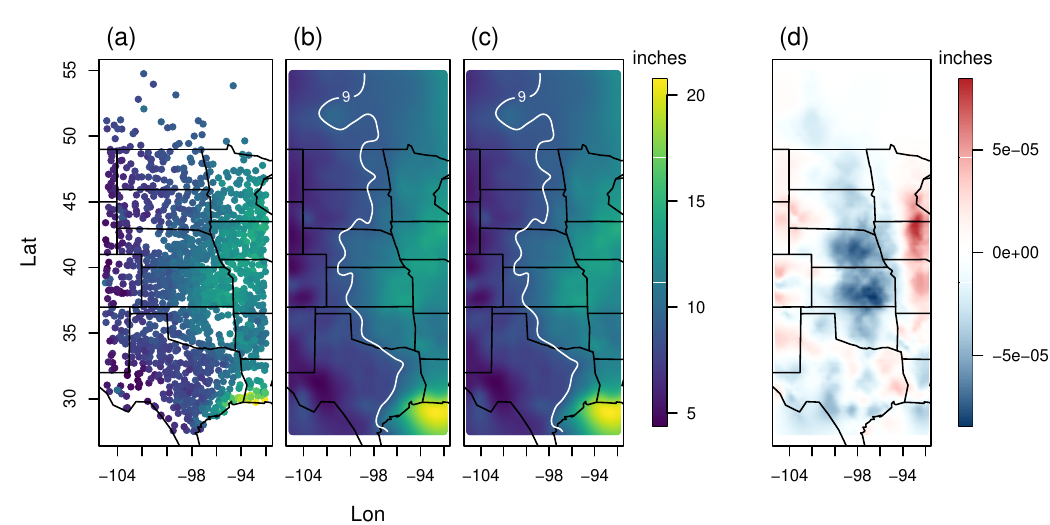}
}
\vspace{-0.4in}
\caption{\label{fig:RainfallPredictionComparison} These plots compare the exact prediction and rapid approximate prediction surface for precipitation in inches for summer rainfall (June, July, August); in white is the predicted contour at 9 inches. (a) the $n = 1368$ data points for this example; (b) exact prediction result onto the prediction grid of size $256 \times 512$ with a contour line at 9 inches; (c) rapid approximate prediction; (d) the point-wise  error between the two predicted surfaces from (b) and (c).}
\end{figure}

From Figure \ref{fig:RainfallPredictionComparison} (b) and (c), we observe that the exact prediction method and rapid prediction method yield visually non-distinguishable results. When we examine the point-wise error between the two methods in (d), we see that the rapid method is accurate to about 4 decimal places. Higher errors are observed where data locations are concentrated, and are lowest where data are sparse. This makes sense since the approximation of kernels centered at data points is the largest at the grid points near this location and improves farther away.

\clearpage

For making spatial inferences about this surface we generate a 100-member ensemble by conditional simulation. We switched to a $64 \times 128$ grid to facilitate the exact computation of the prediction standard error. For comparison we use the approximate CS scheme with local simulation size of 5 and the rapid prediction algorithm with the nearest neighbor order of $L = 4$.

\begin{figure}
\centering{
\includegraphics[width=\textwidth]{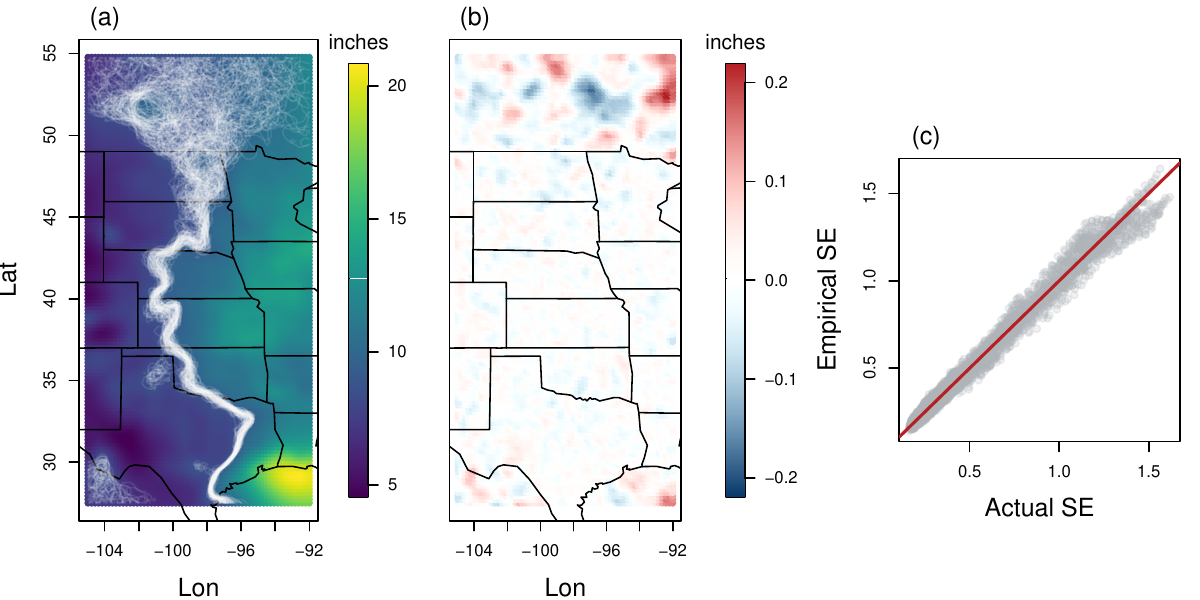}
}
\vspace{-0.4in}
\caption{\label{fig:RainfallCS9inContour} Results from a 100-member ensemble of the fast CS: (a) the mean of the ensemble surface with 9 in contours from the individual ensemble members in gray; (b) the point-wise difference between standard error computed exactly and the empirical one estimated from the ensemble; and (c) shows the empirical and exact standard errors plotted against each other in light gray and a red line representing exact agreement.}
\end{figure}

From Figure \ref{fig:RainfallCS9inContour} (b), we see that the majority of disagreement between the exact Kriging standard error and the empirical standard error are in regions where data are sparse and from (c) we see that there's strong agreement between the empirical and exact standard errors. Thus in this example our fast CS scheme with rapid prediction successfully approximated the actual Kriging standard errors.

\subsection{Timing}\label{sec:RainfallTiming}
Finally we report the timing over several different cases and models. We fit a model under the Vecchia approximation framework, then predict onto evenly spaced grids using functions from the \texttt{GpGp} R package. We do the same with the defaults in the {\tt LatticeKrig} function from the \texttt{LatticeKrig} R package. In the case of LatticeKrig the default is structured to approximate a thin-plate spline with smoothness $m=2$, a second derivative penalty function. In all cases the timing omits the time for covariance parameter estimation and more about this aspect is included in the Discussion section. Each timing sample point is taken from the median of 20 timing runs.

\begin{figure}
\centering{
\includegraphics[width=\textwidth]{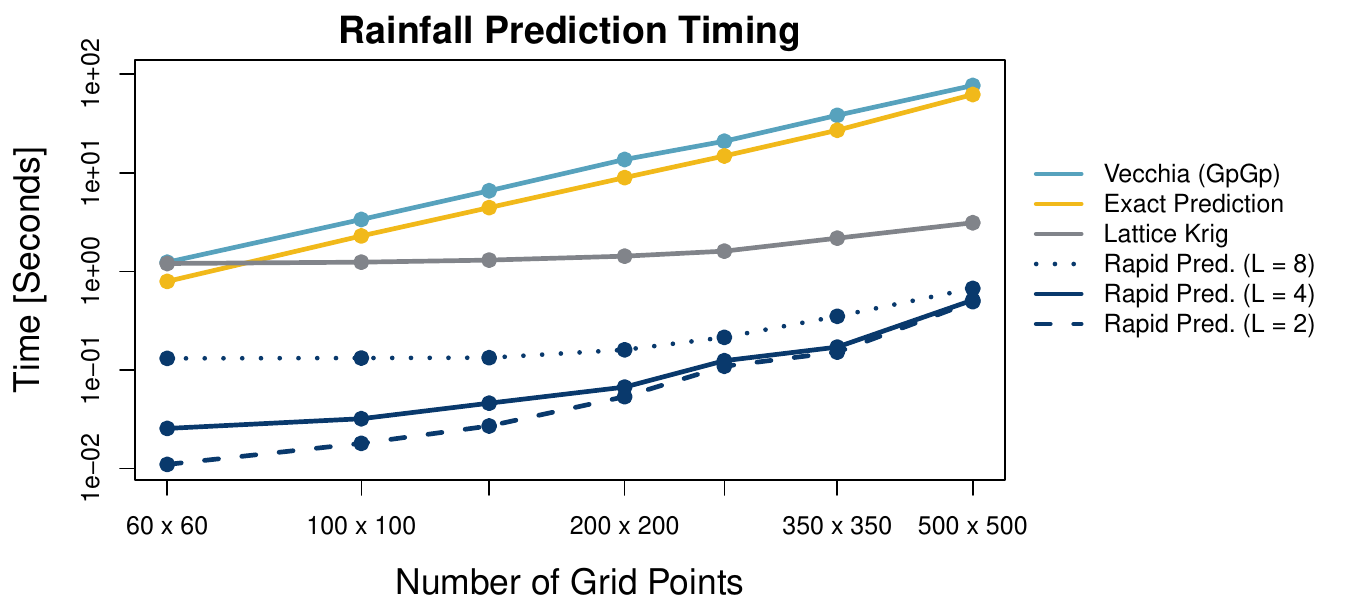}
}
\vspace{-0.3in}
\caption{\label{fig:RainfallMethodTiming} Timing for prediction using rainfall data along the meridian. At the prediction grid size of $350 \times 350$, rapid prediction with the nearest neighbor size of 4 took $0.1705$ seconds, while LatticeKrig took $\approx 2$ seconds, Vecchia approximation took $\approx 38$ seconds, and exact prediction took $\approx 27$ seconds. The rapid method provides a factor of 150 times speedup compared to the exact method at this grid resolution.}
\end{figure}

From Figure \ref{fig:RainfallMethodTiming}, we see that the rapid method at nearest neighbor sizes of 2, 4, and 8 are all faster than the Vecchia approximation, exact method, and {\tt LatticeKrig} for grid sizes larger than $100\times 100$. Although rapid prediction with smaller nearest neighbor sizes is faster at smaller numbers of grid points, this difference in speed shrinks when the number of grid points for prediction becomes large. We believe that at smaller numbers of grid points, the cost of inverting $K_{\mathcal{N}}$ in the algorithm dominates speed while at large numbers of grid points the time is dominated by the FFT.

In the timing experiment for conditional simulation, we generate 10-member ensembles from CS with exact prediction, from fast CS with exact prediction, and from fast CS with rapid prediction using nearest neighbor order $L = 4$. We will also compare these methods to the Vecchia approximation approach in the \texttt{GpGp} package.

\begin{figure}
\centering{
\includegraphics[width=\textwidth]{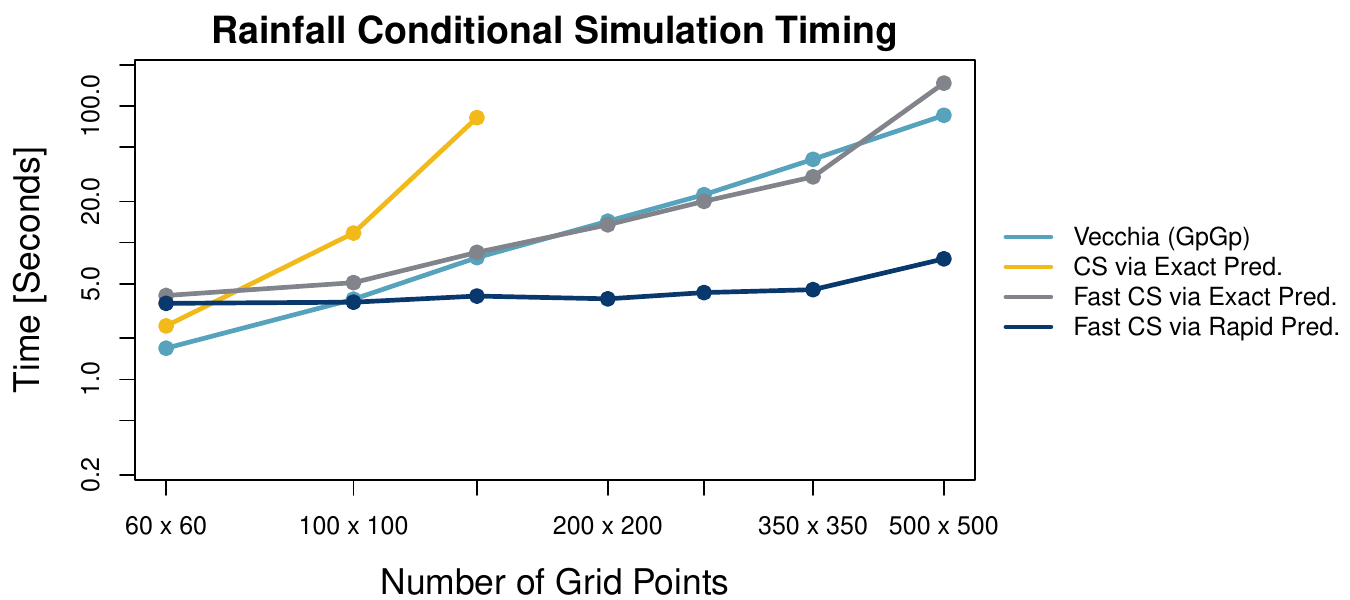}
}
\vspace{-0.3in}
\caption{\label{fig:RainfallCSTiming} Timing for a 10-member ensemble from different conditional simulation schemes using summer rainfall data along the meridian. Timing is done on an Apple Laptop with M2 processor using BLAS, and each data point is taken from the median of 10 timing runs. Timing for conditional simulation via exact prediction stopped at the grid size of $140 \times 140 $ due to excessive run time.}
\end{figure}

Figure \ref{fig:RainfallCSTiming}, reports the timing results for these data.
 The set-up time for rapid prediction is offset by the FFT efficiency starting at $100 \times 100$ grid points.
 The fast CS via rapid prediction stayed around 4 seconds to produce a 10-member ensemble, while other methods continued to increase in time. At the largest grid $500 \times 500$ we see about a factor of 10 speedup over the Vecchia approach.
 
\section{Conclusion \& Discussion}
By approximating a stationary covariance function onto a regular grid we are able to approximate the prediction step in a spatial prediction using convolution via the fast Fourier transform. The net result is a speedup that can easily be a factor of 100 for typical problems encountered in spatial data analysis. Moreover, the approximation also achieves a level of accuracy where statistical inference is not influenced. For example, we have shown that the rapid prediction method provided more than 150 times improvement in speed at $350 \times 350$ grid points ($L=4$) and achieves at least 4 digits of accuracy when compared to the prediction timing using an exact method. Besides just spatial prediction this algorithm is a useful improvement in CS where repeated prediction is needed and again we see dramatic speedup with acceptable accuracy. 

The timing results for CS come with a caveat that one must be able to simulate the unconditional spatial process on a fine grid efficiently. In this work we adopt circulant embedding as the algorithm of choice. However, circulant embedding is not always valid and breaks down for some covariances, typically when the correlation range is similar in size to the extent of the spatial domain. Adapting circulant embedding or devising other fast simulation algorithms for these cases is still an open problem but distinct from our concern in this work and any fast simulation method will benefit from rapid prediction.

The key to the success of this method is using the same fitted kernel at grid locations to approximate the ones that are centered at off-grid observation locations. This enables the use of the FFT to speed up the computation and as expected the overhead with larger grid sizes is small because of the computational order of the FFT with $N$. Besides verifying the accuracy of the approximation numerically in Section 4 we also provide some analytical error bounds from kernel interpolation theory. This confirms the improvement in accuracy as smoothness increases and the grid spacing decreases. Surprisingly the approximation tends to have a better rate (steeper slope) than the theory but this is a feature that has been observed elsewhere when considering interpolation error bounds. We found this analysis useful in giving a foundation for the accuracy without the need for numerical experiments. Moreover the error bounds are not heavily dependent on the form of the covariance function (kernel) and so suggest these results will hold for any covariance with a suitable degree of smoothness. Finally it is of interest to trace the connection between this active area in applied mathematics and spatial statistics as a way to understand more about the case of spatial interpolation.

In this work we have only considered prediction onto a regular grid. What about prediction at a large number of irregular locations?An intermediate step of interpolation accuracy often involves a local Taylor series expansion of the true function. This representation suggests that for fill distances that are small,  local polynomials may be effective in interpolating to off-grid predictions. One simple approach is to take advantage of the current method and prediction onto a fine mesh. Predicted values off of this mesh can be approximated using a simple, fast and parallel interpolation, for example using local polynomials.

A key assumption in the rapid algorithm is that of a stationary covariance function. However, large spatial data, particularly from environmental problems, are often non-stationary and so an extension of the rapid prediction algorithm is an issue. The details of the algorithm suggest how this might be done. At Step 3 the covariance vector $\mathbf{k}_i$ is found between the observation location, $i$, and the neighboring grid points. As written this is the same covariance as is used for the approximation. However, the stationary covariance used then in Step 4 to interpolate $\mathbf{k}_i$ need not be the same as the nonstationary covariance. If the nonstationary covariance has smoothness and scaling that is similar to some ``average" stationary covariance used for the interpolation, the approximation will be accurate and our method will be useful. Note that in terms of the timing we would expect the speedup to be the same. The only difference is in Step 3 to evaluate the non-stationary covariance and then subsequent steps to find $A_i$ are the same amount of computation. Another approach is to consider a nonstationary covariance that is a weighted combination of several stationary covariances but where the weights vary over space. In this case one would use rapid prediction to evaluate each of the stationary components and then combine them together using the spatially dependent weight functions. This second strategy is well suited to spatial processes that have a multi-resolution structure where the spatial process can have very different correlation scales at different locations in the spatial domain. In either of these cases the key features of the algorithm are unchanged and so our current development for stationary covariance prediction is valuable as the foundation for algorithmic extensions.

Although the Vecchia approximation was not competitive in this work focused on prediction it has an important strength in parameter estimation. To find maximum likelihood estimates for the covariance parameters we, of
course, cannot exploit the rapid prediction algorithm. For the rainfall example finding the MLE using the exact
likelihood took more than two minutes and this is in contrast to a time of 2 seconds using the Vecchia
approximation. Moreover, the form for the Vecchia covariance guarantees that the same covariance parameters are being estimated with this approximation so it carries guarantees for parameter consistency. Thus a hybrid strategy is to use a different method, such as the Vecchia approximation, to estimate parameters for the covariance function and then use the rapid prediction method to perform predictions at fine grids. The setup time for rapid prediction is small and so using different sets of covariance parameters for each predicted field is feasible. This makes the algorithm amenable to bootstrapping or conditional sampling where the covariance parameters change for each sample. In general we believe that pursuing hybrid numerical approximations for well-posed spatial models to facilitate fast and possibly interactive data analysis is a fruitful area. In particular fast operations that rely on structured grids can be valuable for efficient computations not only for frequentist method but as a component of a full Bayesian analysis. 

\section{Acknowledgments}\label{acknowledgments} 
We thank Dr. Joseph Guinness for verifying the correct implementation of the Vecchia approximation in our timing experiments. 

\section{Disclosure statement}\label{disclosure-statement}

The authors have no conflict of interest regarding this work. 

\section{Data Availability Statement}\label{data-availability-statement}

Relevant experiment data and code are available at the following URL: \url{https://github.com/ziyuli22/Rapid_Approximation_Method_2026}. 

\appendix
\section{Appendix: Tables}
\label{appendix:a}
\renewcommand{\thetable}{A\arabic{table}}
\setcounter{table}{0}

\begin{longtable}[]{@{}llll@{}}
\caption{ANOVA results for $\log$ of absolute error of the center evaluation point from the MC experiment using all factors and their interactions.}\label{tab:fullANOVA}\tabularnewline
\toprule\noalign{}
Factor & Degrees of Freedom &  Mean Square \\
\midrule\noalign{}
\endfirsthead
\toprule\noalign{}
Factor & Degrees of Freedom &  Mean Square \\
\midrule\noalign{}
\endhead
\bottomrule\noalign{}
\endlastfoot
corrDist            & 2 &  6.50  \\
smoothness              & 2 &  192.22  \\
nugget                  & 2 &   30.74  \\
NN                      & 2 & 367.59 \\
nGrid                   & 2 &   32.38  \\
nObs                    & 2 &    13.90  \\
corrDist:smoothness & 4 &    0.19  \\
corrDist:nugget     & 4 &     0.06  \\
corrDist:NN         & 4 &     4.09  \\
corrDist:nGrid      & 4 &    2.00 \\
corrDist:nObs       & 4 &     0.01 \\
smoothness:nugget       & 4 &     0.25  \\
smoothness:NN           & 4 &   31.63 \\
smoothness:nGrid        & 4 &    0.62  \\
smoothness:nObs         & 4 &     0.01 \\
nugget:NN               & 4 &      0.00  \\
nugget:nGrid            & 4 &    0.00  \\
nugget:nObs             & 4 &     0.04 \\
NN:nGrid                & 4 &    5.09 \\
NN:nObs                 & 4 &      0.00 \\
nGrid:nObs              & 4 &     0.00  \\
\end{longtable}

\begin{longtable}[]{@{}llll@{}}
\caption{Regression coefficients for the smaller linear model fit to $\log$ of average absolute error from the MC experiment.}\label{tab:reducedLM}\tabularnewline
\toprule\noalign{}
 & Coef. Estimate & Standard Error  \\
\midrule\noalign{}
\endfirsthead
\toprule\noalign{}
 & Coef. Estimate & Standard Error  \\
\midrule\noalign{}
\endhead
\bottomrule\noalign{}
\endlastfoot
(Intercept)        & -2.36678 & 0.05965 \\
smoothness1        & -0.79492 & 0.07019 \\
smoothness1.5      & -1.52717 & 0.07019 \\
NN4                & -1.08912 & 0.07019 \\
NN8                & -1.80025 & 0.07019 \\
nGrid350           & -0.57619 & 0.04053  \\
nGrid500           & -0.67630 & 0.04053 \\
nugget0.1          & -0.39903 & 0.04053 \\
nugget0.5          & -0.70952 & 0.04053 \\
smoothness1:NN4    & -0.44811 & 0.09927 \\
smoothness1.5:NN4  &  0.20004 & 0.09927  \\
smoothness1:NN8    & -2.04626 & 0.09927 \\
smoothness1.5:NN8  &  0.07016 & 0.09927\\
\end{longtable}

\section{Appendix: Derivation of Error Bound}
\label{appendix:b}
Definition of reproducing kernel from \citep[Definition 13.1]{Fasshauer2007}, adapted for our notation, states that: 
\begin{definition}\label{def:RK}
Let  $\mathcal{H}$ be a Hilbert space of functions $f : \mathbb{R}^d \rightarrow \mathbb{R}$ with norm $\| \cdot \| $ and inner product $\langle \cdot, \cdot \rangle$. A function $\Phi: \mathbb{R}^d \times \mathbb{R}^d \rightarrow \mathbb{R}$ is called the \textbf{reproducing kernel} for $\mathcal{H}$ if
	\begin{enumerate}
	\item $\Phi(\cdot, \mathbf{x}) \in \mathcal{H}$ for all $\mathbf{x} \in \mathbb{R}^d$,
	\item $f(\mathbf{x}) = \langle f, \Phi(\cdot, \mathbf{x}) \rangle$ for all $f \in \mathcal{H}$ and all $\mathbf{x} \in \mathbb{R}^d$.
	\end{enumerate}
\end{definition}

Both conditions from Definition~\ref{def:RK} are satisfied by positive definite covariance functions such as the Mat\'{e}rn covariance family. Notably, condition (2) here is also known as the \textit{reproducing property}. 

We are interested in quantifying error between the interpolant $ \hat{h} \in \mathcal{H}$ and unknown function $h$ where we obtained observations $\{ (\mathbf{x}_i , \mathbf{y}_i )\}_{i=1}^n$ from. We can use \citep[Theorem 14.2]{Fasshauer2007} which, adapted to our notation, states that:

\begin{theorem}\label{thm:PowerFunctionBound}
Let $\Omega \subseteq \mathbb{R}^d$, $\Phi \in C(\Omega \times \Omega)$ be strictly positive definite on $\mathbb{R}^d$, and suppose the points $\mathcal{X} = \{\mathbf{x}_i\}_{i=1}^n$ are distinct. Denote the interpolant to $h \in \mathcal{N}_\Phi (\Omega)$ by $\hat{h}$, then for every $\mathbf{x} \in \Omega$
$$\vert h(\mathbf{x}) - \hat{h}(\mathbf{x}) \vert \leq \mathbf{P}_\Phi (\mathbf{x}) \Vert h \Vert_{\mathcal{N}_\Phi (\Omega)}.$$
Here, $\mathcal{N}_\Phi(\Omega)$ is the native space. 
\end{theorem}

In this theorem, the native space is defined as the reproducing kernel Hilbert space $\mathcal{H}$ induced by a positive definite kernel $\Phi$. This space depends on the kernel, $\Omega$, and $\mathcal{X}$. For ease of notation, we will just use $\mathcal{H}$ to denote this space and rewrite the approximation bound as
\[  | h(\mathbf{x}) - \hat{h}(\mathbf{x}) |   \le  \mathbf{P}_\Phi (\mathbf{x})  \| h \|_{\cal H} .  \]

Note that this bound is point-wise. 

We can further bound the power function by a measure called the fill distance $\delta$ \citep[Equation 2.3]{Fasshauer2007} which, adapted to our notation, is defined as
$$\delta = \delta_{\mathcal{X}, \Omega} = \sup_{\mathbf{x} \in \Omega} \min_{\mathbf{x}_i \in \mathcal{X}} \Vert \mathbf{x} - \mathbf{x}_i \Vert_2.$$
This quantity denotes the largest empty ball that can be placed among the data locations and is a measure of how well data fills out the domain.  
Then, \citep[Theorem 14.5]{Fasshauer2007} states that:

\begin{theorem}\label{thm:FillDistanceBound}
Suppose $\Omega \subset \mathbb{R}^d$ is bounded and satisfies an interior cone condition. Suppose $\Phi \in C^{2\kappa} (\Omega \times \Omega)$ is symmetric and strictly positive definite. Denote the interpolant to $h \in \mathcal{N}_\Phi(\Omega)$ on the set $\mathcal{X}$ by $\hat{h}$. Then there exist positive constants $\delta_0$ and $C_1$ (independent of $\mathbf{x}$, $h$ and $\Phi$) such that
$$\vert h(\mathbf{x}) - \hat{h}(\mathbf{x}) \vert \leq C_1 \delta^\kappa_{\mathcal{X}, \Omega} \sqrt{C_\Phi(\mathbf{x})} \Vert h \Vert_{\mathcal{N}_\Phi (\Omega)}$$
provided $\delta_{\mathcal{X}, \Omega} \leq \delta_0$. Here
$$C_\Phi (\mathbf{x}) = \max_{|\beta| = 2k} \max_{\mathbf{w}, \mathbf{z} \in \Omega \and B(\mathbf{x}, c_2\delta_{\mathcal{X}, \Omega})} \left| D^\beta_2 \Phi(\mathbf{w}, \mathbf{z}) \right|$$
with $B(\mathbf{x}, c_2 \delta_{\mathcal{X}, \Omega})$ denoting the ball of radius $c_2 \delta_{\mathcal{X}, \Omega}$ centered at $\mathbf{x}$. 
\end{theorem}

The domain $\Omega$ in almost all spatial statistics applications do satisfy the interior cone condition. And from this theorem, we know that $C_\Phi(\mathbf{x})$ is bounded for a kernel like the Mat\'{e}rn that has finite smoothness and on $\Omega$ bounded.  So suppose that we denote the positive constant $C(\mathbf{x}) = C_1 \sqrt{C_\Phi(\mathbf{x})}$ and the fill distance $\delta$ is less than some critical fill distance $\delta_0$, then this bound is 
$$\vert h(\mathbf{x}) - \hat{h}(\mathbf{x}) \vert \leq C(\mathbf{x}) \delta^\kappa\Vert h \Vert_{\mathcal{H}}.$$
Taking the supremum, we get 
\begin{equation}
\sup_{\mathbf{x} \in \Omega} \left| h(\mathbf{x}) - \hat{h}(\mathbf{x}) \right| \le C \delta^\kappa \tag{\ref{eq:supKernelBound}}
\end{equation}
for $C = \sup_{\mathbf{x} \in \Omega} C(\mathbf{x}) \Vert h \Vert_\mathcal{H}$.

\clearpage





  \bibliography{bibliography.bib}

\end{document}